\documentclass[12pt]{article}
\usepackage{amsmath}
\usepackage{caption}
\usepackage{calc}
\usepackage{graphicx}
\usepackage{multicol}
\usepackage{cite}
\usepackage{color}

\makeatletter
\renewcommand{\maketitle}{\bgroup\setlength{\parindent}{0pt}
	\begin{flushleft}
		\textbf{\@title}
		
		\@author
	\end{flushleft}\egroup
}
\makeatother
\begin{document}

\title{\bf \hspace{1cm} {Bipolar Expansions and Overlap  Corrections to the \\   \hspace{1cm} Electrostatic Interaction Energy\\ \vspace{1cm}}}

\author{ \hspace{1cm}{ G.Vaman}   \\  \hspace{1cm} Institute of Atomic Physics, P.O.Box MG-6, Bucharest, 
	Romania, \\\hspace{1cm} vaman@ifin.nipne.ro}
\maketitle
\abstract
	
\noindent{\it {\bf Abstract} We use the multipole technique to derive four equivalent expressions for the bipolar expansion of the inverse distance, valid in all the regions of configuration space. Using the first-order perturbation theory, we calculate the overlap correction to the 
long-range electrostatic energy between two hydrogen atoms and between a hydrogen atom and a proton. }\vspace{0.5cm}

\begin{multicols}{2}
\section*{\small{1 Introduction}}

The   bipolar expansion of the inverse distance between two points and its applications in the study of chemical bonding has been widely studied in the literature by different methods \cite{bue,sac,kay}, as being very suitable for approximating the interaction potential between various kinds of molecules in terms of their multipoles. It is usually used in a four-region form, which  imply the decomposition of the integration space in regions with complicated boundary-surfaces: a nonoverlapping region and three overlapping regions. As many modern numerical calculations used in quantum chemistry are based on multipolar approximations \cite{toi, bec}, the convergence of the bipolar expansion is very important.    
 The recent conclusions of the numerical  studies from Refs. \cite{sil1, sil2} are not encouraging at all. They reveal that the convergence in the overlapping region is  conditional and that the Laplace equation is not satisfied either termwise or pointwise. Nevertheless, the author of \cite{sil1} hopes in "the possibility of clarification through generalized functions".

 In this paper, we are going to   
 analyze  new forms of the bipolar expansion, deduced from the multipolar formalism of Refs. \cite{dub,rad} and  calculate the overlap correction to the long-range electrostatic energy for two simple systems: a hydrogen atom and a proton and two hydrogen atoms in their ground states.
Using this formalism we obtain well-known results \cite{pau} and this encourages us to assert that our expansions are "good enough" for studying more complicated systems. 
In contrast to other papers which are studying the bipolar expansions, we start from the multipole series of the electrostatic interaction energy and compare it with its definition as can be found in textbooks.  Our results contain well-behaved functions as well as generalized functions, therefore their convergence can be discussed only in the weak or distributional sense. In our approach, using the results of Ref.\cite{dub}, we are able to identify the generalized-functions terms with the contribution of the mean square radii and to prove that they describe the overlap (or contact) interaction. \\
The paper is organized as follows:
in the next section we justify four equivalent forms for the bipolar expansion of the inverse distance, valid in all regions of space. 
Next,  we use the perturbation theory to calculate the first-order correction to the electrostatic interaction energy for two simple physical systems, neglecting the exchange. The end section is devoted to conclusions.

\section*{\small{2 New Forms of the Bipolar Expansion}}
We consider two charge distributions, one of which is centered at the origin and is described by the vector ${\bf b}$ and the other centered at a distance $R$ from the origin and described by the vector ${\bf a}$, as shown in Fig. 1.
We are concerned with the inverse distance between the points A and B.
We shall  justify here the following four equivalent equations:

\end{multicols}
\begin{figure}[h]
	\centering
	\includegraphics[width=70mm]{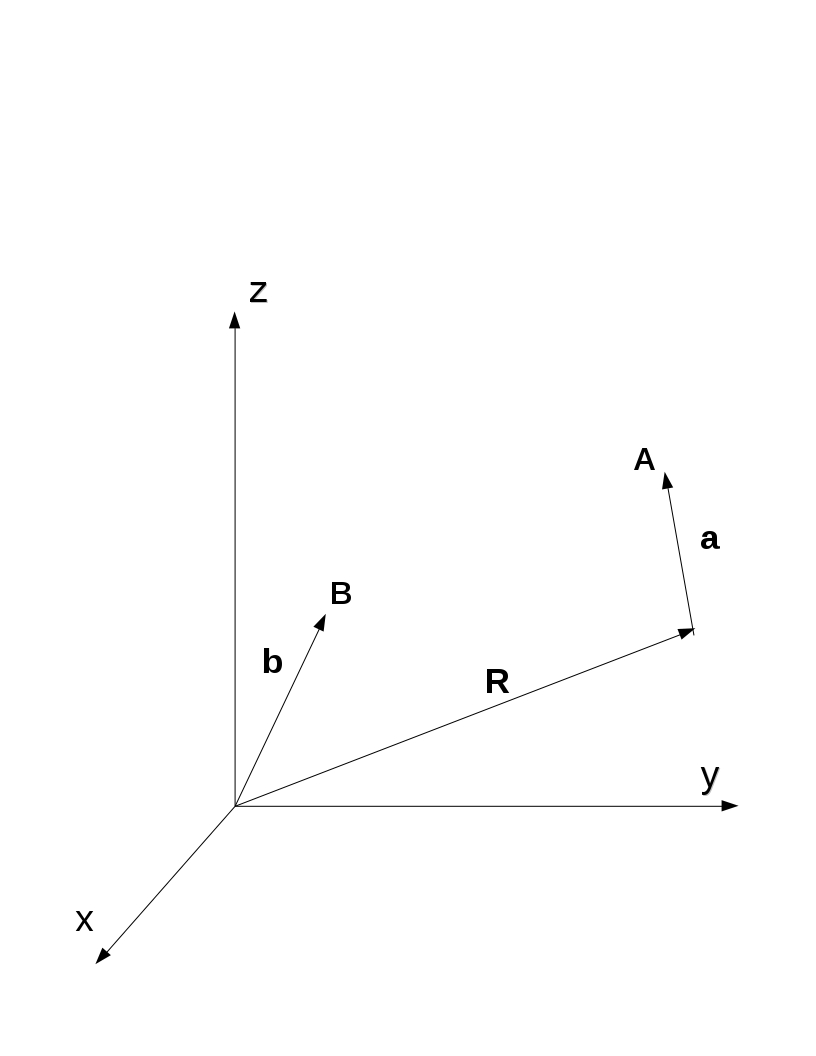}
	\caption{Coordinates for the bipolar expansion}
	
	\label{fig:1}       
\end{figure}

\begin{align}\label{sf0}
	\frac{1}{| {\bf b}-{\bf a}- {\bf R} | } = \hspace{11.5cm} \nonumber \\
	\sum_{l=0}^{\infty} \sum_{l'=0}^{\infty}
	\sum_{m=-l}^{l} \sum_{m'= -l'}^{l'} \sum_{n =0}^{\infty} \sum_{n' =0}^{\infty}  \frac{ (-1)^{l'} (4 \pi)^2  b^{l+2n} a^{l'+2n'} }{2^{n+n'} n! n'! (2l+2n+1)!! (2l'+2n'+1)!!} \cdot \hspace{3cm} \\
	Y^*_{lm}( {\hat{\bf b}}) Y^*_{l'm'}( \hat{{\bf a}}) \Delta_{\bf R}^{n+n'} Y_{lm}(- \nabla_{\bf R}) Y_{l'm'}(- \nabla_{\bf R}) \frac{1}{R}.\hspace{3cm} \nonumber
\end{align}

\begin{align} \label{sf}
	\frac{1}{| {\bf b}-{\bf a}- {\bf R} | } = \hspace{12cm} \\ \sum_{l=0}^{\infty} \sum_{l'=0}^{\infty}
	\sum_{m=-l}^{l} \sum_{m'= -l'}^{l'} \sum_{n =0}^{\infty} \sum_{n' =0}^{\infty}\frac{ (-1)^{l} (4 \pi)^2  b^{l+2n} a^{l'+2n'} \langle l+l', m+m'|lm|l'm'\rangle}{2^{n+n'} n! n'! (2l+2n+1)!! (2l'+2n'+1)!!} \cdot \hspace{3cm} \nonumber\\
	Y^*_{lm}(\hat{{\bf b}}) 	Y^*_{l'm'}( \hat{{\bf a}})
	\Delta_{ {\bf R}}^{n+n'}  Y_{l+l'\; m+m'} ( \nabla_{\bf R})\frac{1}{R}, \hspace{3cm} \nonumber 
\end{align}

\begin{align} \label{car}
	\frac{1}{| {\bf b}-{\bf a}- {\bf R} | }= \hspace{12cm} \\
	\sum_{l=0}^{\infty} \sum_{l'=0}^{\infty}
	\sum_{m=-l}^{l} \sum_{m'= -l'}^{l'} \sum_{n =0}^{\infty} \sum_{n' =0}^{\infty}  
	\frac{(-1)^{l'} (2l+1) (2l'+1)  b^{2l+2n+1} a^{2l'+2n'+1}}{2^{n+n'}n!\; n'!\; l! \;l'!\; (2l+2n+1)!! 
		(2l'+2n'+1)!!} \cdot\nonumber \hspace{3cm} \\
	\partial_{{\bf b} i_1} \dots \partial_{{\bf b} i_l} \frac{1}{b} \partial_{{\bf a} j_1} \dots \partial_{{\bf a} j_l'} \frac{1}{a} \Delta_{\bf R}^{n+n'} 
	\partial_{{\bf R} i_1} \dots \partial_{{\bf R} i_l}  \partial_{{\bf R} j_1} \dots \partial_{{\bf R} j_l'} \frac{1}{R}, \hspace{3cm} \nonumber
\end{align}

\begin{align} \label{carcon}
	\frac{1}{| {\bf b}-{\bf a}-{\bf R} | }= &\sum_{l=0}^{\infty} \sum_{l'=0}^{\infty}
	\sum_{n =0}^{\infty} \sum_{n' =0}^{\infty} \sum_{\begin{array}{c}
			p_x, p_y, p_z \\
			p_x+p_y+p_z=l
		\end{array}}
		\sum_{\begin{array}{c}
				q_x, q_y, q_z \\
				q_x+q_y+q_z=l
			\end{array}}
			\nonumber \\
			&\frac{(-1)^{l'} (2l+1) (2l'+1)  b^{2l+2n+1} a^{2l'+2n'+1}}{2^{n+n'}n! n'!  (2l+2n+1)!! 
				(2l'+2n'+1)!!p_x! p_y! p_z! q_x! q_y! q_z!} \cdot \nonumber \\
			& \partial_{b_x}^{p_x} \partial_{b_y}^{p_y} \partial_{b_z}^{p_z} \frac{1}{b} 
			\partial_{a_x}^{q_x} \partial_{a_y}^{q_y} \partial_{a_z}^{q_z} \frac{1}{a}
			\Delta_{\bf R}^{n+n'} 
			\partial_{R_x}^{p_x+q_x} \partial_{R_y}^{p_y+q_y} \partial_{R_z}^{p_z+q_z} \frac{1}{R} , 
		\end{align}
		\begin{multicols}{2}
		where we have used the following notations:
		\begin{itemize}
			\item $r_i$ is the $i$-th component of the vector ${\bf r}$ and $ r = | {\bf r}|$, , $\hat{\bf r}= {\bf r}/r$
			\item[-]$Y_{lm}(\hat{\bf r})$ is a spherical harmonic and $Y_{l m}( {\bf r})=r^l Y_{lm}(\hat{\bf r})$ is a solid spherical harmonic (\cite{app})
			\item[-] $ \partial_{{\bf r}i}= \frac{\partial}{\partial {\bf r}_i} $ means partial derivative with respect to the $i$-th component of the vector ${\bf r}$
			\item[-]$\nabla$ is the gradient operator and $\Delta = \nabla \cdot \nabla$ is the laplacian
			\item[-] $\partial_{{\bf r}}^n= \underbrace{\partial_{{\bf r}} \dots    \partial_{{\bf r}} }_{ \text{n times}} $, 
			$\Delta_{{\bf r}}^n= \underbrace{\Delta_{{\bf r}} \dots    \Delta_{{\bf r}} }_{ \text{n times}} $
			\item $\langle l+l', m+m'|lm|l'm'\rangle $ is a Gaunt coefficient \cite{wen}.
		\end{itemize}
		 The summation over dummy indices is understood.
		In the following, for shortness,  we shall habitually drop the limits in summations.

		According to Ref. \cite{dub}, the two charge distributions  are described by the expansions:
		\end{multicols}
		\begin{align} \label{dens1}
			\rho_1( {\bf r} ) = \sum_{l, m, n} \frac{(2l+1)!!}{2^n n! (2l+2n+1)!!} \sqrt{ \frac{4 \pi}{2l+1}} \overline{r_{lm}^{2n}}^{(1)} \Delta^n \delta_{lm}({\bf r}),
		\end{align}

		\begin{align} \label{dens2}
			\rho_2( {\bf r} ) = \sum_{l, m, n} \frac{(2l+1)!!}{2^n n! (2l+2n+1)!!} \sqrt{ \frac{4 \pi}{2l+1}} \overline{r_{lm}^{2n}}^{(2)} \Delta^n \delta_{lm}({\bf r} - {\bf R}),
		\end{align}
		\begin{multicols}{2}
	\noindent	where $	\rho( {\bf r} ) $ is the charge density, $\delta({\bf r})$ is the Dirac delta function, $\delta_{lm}({\bf r})= \frac{1}{(2l-1)!!} Y_{lm}(-\Delta) \delta({\bf r})  $
	and	 the two sets of  mean square radii $\overline{r_{lm}^{2n}}^{(1,2)}$ are calculated with respect to the two centers of the charge distributions:
	
		\begin{align}\label{raze}
			\overline{r_{lm}^{2n}}^{(1)} = 
			\sqrt{ \frac{4 \pi}{2l+1}} \int b^{l+2n} Y_{lm}^*( \hat{\bf b}) \rho_1({\bf b}) d^3 b,\nonumber\\ 
			\overline{r_{lm}^{2n}}^{(2)} = 
			\sqrt{ \frac{4 \pi}{2l+1}} \int a^{l+2n} Y_{lm}^*( \hat{\bf a}) \rho_1({\bf a}) d^3a.
		\end{align}
	
		The electrostatic potential produced by the second charge distribution is:
		
		\begin{align} \label{potential}
			\phi_2({\bf r}) = \sum_{l,m,n} \frac{(2l+1)}{2^n n! (2l+2n+1)!!} \sqrt{ \frac{4\pi}{2l+1}} \cdot \nonumber \\ \overline{r_{lm}^{2n}}^{(2)} \Delta^n Y_{lm}(- \nabla
			) \frac{1}{| {\bf r} - {\bf R}|}.
		\end{align}
	
		After integrating by parts, one obtains the following expression for the electrostatic interaction energy
		\begin{align} \label{enpot}
			 W= \int d^3r \rho_1({\bf r}) \phi_2( {\bf r}):
			 \end{align}
		\end{multicols} 
		\begin{align}\label{ensf}
			W=& \sum_{l, l', m, m', n, n'} \frac{(-1)^{l'} 4\pi  \sqrt{(2l+1) (2l'+1)}  }{ 2^{n+n'} n! n'! (2l+2n+1)!! (2l'+2n'+1)!!}  \overline{r_{lm}^{2n}}^{(1)} \overline{r_{l'm'}^{2n'}}^{(2)} \cdot \nonumber \\
			& \Delta_{\bf R}^{n+n'} Y_{lm} (- \nabla_{\bf R}) Y_{l'm'} (- \nabla_{\bf R}) \frac{1}{R}.
		\end{align}
		\begin{multicols}{2}
		We insert in the above equation the expressions of the mean square radii (\ref{raze}) and interchange sums and integrals:
		\end{multicols}
		\begin{align}
			W=&\int d{\bf b} \int d{\bf a} \;\rho_1({\bf b}) \; \rho_2({\bf a}) \sum_{l, l', m, m', n, n'} \frac{ (-1)^{l'} (4\pi)^2 b^{l+2n} a^{l'+2n'}}{2^{n+n'} n! n'! (2l+2n+1)!! (2l'+2n'+1)!!} \cdot \nonumber \\
			&Y_{lm}^* ( \hat{{\bf b}}) Y_{l'm'}^* ( \hat{{\bf a}}) \Delta_{\bf R}^{n+n'} Y_{lm} (- \nabla_{\bf R}) Y_{l'm'} (- \nabla_{\bf R}) \frac{1}{R}.
		\end{align}
		\begin{multicols}{2}
		If we compare the above equation with:
		\begin{align} \label{encon}
			W=&\int d{\bf b} \int d{\bf a} \frac{ \rho_1({\bf b}) \; \rho_2({\bf a})}{ | {\bf b}- {\bf a}- {\bf R}|},
		\end{align}
		we easily obtain the expansion (\ref{sf0}). Eq.(\ref{sf0}) can be easily verified in the Fourier space: if we take the Fourier transform $\int d {\bf R} \; e^{i {\bf k} {\bf R}}$ and make the summations over the indices $n$ and $n'$ we obtain:
		\end{multicols}
		\begin{align}
			e^{i {\bf k}({\bf b}-{\bf a})} = \sum_{l,l',m,m'} j_l(kb) j^*_{l'}(ka)Y_{lm}^* ( \hat{{\bf b}}) Y_{l'm'}^* ( \hat{{\bf a}}) Y_{lm} ( \hat{{\bf k}}) Y_{l'm'} ( \hat{{\bf k}}),
		\end{align}
		\begin{multicols}{2}
		where \[j_l(x) = 4 \pi i^l \sum_{n=0}^{\infty} \frac{ (-1)^n x^{l+2n}}{n! 2^n (2l+2n+1)!!}\]are the spherical Bessel functions. Thus, we have obtained the well-known Rayleigh expansion of the plane wave.
		Eq.(\ref{sf}) is a slightly modified form of Eq.(\ref{sf0}), obtained by using the addition theorem of the irregular solid spherical harmonics  from \cite{wen}, Eq. (7.4), which we transcribe here in a slightly modified form:
		\begin{align}
		\frac{1}{(2l'-1)!!} Y_{lm}(-\nabla_{\bf R})  Y_{l'm'}(-\nabla_{\bf R}) \frac{1}{R} = 
		\nonumber \\
		 \langle l+l' \; m+m'|l m | l' m' \rangle 
		 2^{l} \frac{ \left( \frac{1}{2} \right)_{l+l'}}{\left( \frac{1}{2} \right)_{l'}} \cdot \nonumber \\
		 \frac{1}{(2l+2'-1)!!}Y_{l+l', m+m'}(-\nabla_{\bf R}) \frac{1}{R}.
		\end{align}	
		After we use the definition of the Pochhammer symbol: $(a)_n= \frac{\Gamma(a+n)}{\Gamma(a)}$, where $\Gamma(x)$ is the Euler Gamma function, the above equation becomes:
		\begin{align}
		 Y_{lm}(-\nabla_{\bf R})  Y_{l'm'}(-\nabla_{\bf R}) \frac{1}{R} = \nonumber \\
		\langle l+l' \; m+m'|l m | l' m' \rangle Y_{l+l', m+m'}(-\nabla_{\bf R}) \frac{1}{R}
		\end{align}
		and our Eq.(2) follows immediately.  
		
		Now, in Eq.(\ref{ensf}) we pass to the Cartesian tensor formalism, using Eq.(16) of Ref.\cite{rad}, which can be written in operatorial form as follows:
		\begin{align}
		\sum_m \overline{r_{lm}^{2n}} Y_{lm}(-\nabla) = \nonumber \\
		(2l-1)!! \sqrt{\frac{2l+1}{4\pi}} \frac{(-1)^l}{l!}  \overline{ r_{i_1}^{2n}}_{\dots i_l} \partial_{i_1} \dots \partial_{i_l}
		\end{align} 
			  We obtain the following expression for the electrostatic interaction energy:
		\end{multicols}
		\begin{align}\label{encar}
			W=& \sum_{l,l',m,m',n,n'} \frac{(-1)^l (2l+1)!! (2l'+1)!!}{2^{n+n'} n! n'! l! l'! (2l+2n+1)!! (2l'+2n'+1)!!} \overline{ r_{i_1}^{2n}}^{(1)}_{\dots i_l} \; \overline{ r_{j_1}^{2n'}}^{(2)}_{\dots j_l'} \nonumber \\
			& \Delta_{{\bf R}}^{n+n'} \partial_{i_1} \dots \partial_{i_l} \partial_{j_1} \dots \partial_{j_{l'}} \frac{1}{R},
		\end{align}
		\begin{multicols}{2}
		where \[ \overline{ r_{i_1}^{2n}}_{\dots i_l} = \frac{(-1)^l}{(2l-1)!!} \int d{\bf r} \; r^{2l+2n+1} \rho({\bf r}) \partial_{i_1} \dots \partial_{i_l} \frac{1}{r} \]
		are the Cartesian components of the mean square radius of order $n$ and multipolarity $l$, and the summation over dummy indices is understood.
		If we introduce in Eq.(\ref{encar}) the above expression for the Cartesian mean square radius and compare with Eq.(\ref{encon}), we obtain Eq.(\ref{car}).  Eq.(\ref{carcon})  can be obtained by writing the contractions of the totally symmetric tensors in terms of the compressed forms of the tensors (Eq.(2) of Ref.\cite{app}), as follows:
		\begin{align}
		\partial_{i_1} \dots \partial_{i_l} \partial_{R_{i_1}} \dots \partial_{R_{i_l}}= \nonumber \\\sum_{\begin{array}{c}
			p_x, p_y, p_z \\
			p_x+p_y+p_z=l
			\end{array}} \frac{p!}{p_x! p_y! p_z!} \partial_x^{p_x}  \partial_y^{p_y}  \partial_z^{p_z}  \partial_{R_x}^{p_x}  \partial_{R_y}^{p_y}  \partial_{R_z}^{p_z}.
		\end{align}

		\section*{\small{3 Calculation of the Long-range Electrostatic Energy}}
		\label{sec:3}

		\subsection*{\small{ 3.1 The interaction between two hydrogen atoms}}
		\label{sec:31}
		In this section we consider the long-range electrostatic interaction between two hydrogen atoms, in their ground states. We shall neglect the exchange and calculate the first order correction to the energy by using the non-degenerate perturbation theory. As the charge density of a hydrogen atom centered at the origin is: 
		\[ \rho({\bf r}) = e \delta({\bf r}) - e \delta({\bf r}-{\bf a}),\]
		where $e > 0$ is the charge of the proton, the mean square radii of the two interacting systems (calculated in their own reference systems) are: 
		\begin{align} \overline{r_{lm}^{2n}}^{(1)}= -e \sqrt{\frac{4\pi}{2l+1}} a^{2n} Y_{lm}^* ({\bf a}) +e \delta_{l,0} \delta_{n,0}, \hspace{0.7cm}\nonumber\\ 
			\overline{r_{l'm'}^{2n'}}^{(2)}= -e \sqrt{\frac{4\pi}{2l'+1}} b^{2n'} Y_{l'm'}^*({\bf b}) +e \delta_{l',0} \delta_{n',0}. \nonumber
		\end{align}
		Introducing the above expressions in Eq.(\ref{ensf}), we obtain the following expression of the perturbation operator:
		\end{multicols}
		\begin{align}\label{oper}
			W({\bf a}, {\bf b}; {\bf R})= & \sum_{l,l',m,m', n,n'} \frac{ (4\pi)^2 e^2 (-1)^{l'} a^{2n} b^{2n'}}{2^{n+n'} n! n'! (2l+2n+1)!! (2l'+2n'+1)!!} Y_{lm}^*({\bf a}) Y_{l'm'}^*({\bf b}) \cdot \nonumber \\
			& \Delta_{\bf R}^{n+n'} Y_{lm}(-\nabla_{\bf R}) Y_{l'm'}(-\nabla_{\bf R}) \frac{1}{R}- \nonumber \\
			& \sum_{l,m,n} \frac{4\pi e^2 a^{2n}}{2^n n! (2l+2n+1)!!} Y_{lm}^*({\bf a}) \Delta_{\bf R}^n Y_{lm}(-\nabla_{\bf R}) \frac{1}{R} - \\
			& \sum_{l',m',n'} \frac{(-1)^{l'} 4\pi e^2 b^{2n'}}{2^{n'} n'! (2l'+2n'+1)!!} Y_{l'm'}^*({\bf b}) \Delta_{\bf R}^{n'} Y_{l'm'}(-\nabla_{\bf R}) \frac{1}{R} + \frac{e^2}{R}. \nonumber
		\end{align}
		\begin{multicols}{2}
	\noindent	Note that, if we place the $x$-axis along ${\bf R}$ and explicitly write the first terms of the above expansion, we obtain the r.h.s. of Eq.(3) from \cite{mar}, supplemented by an  infinite number of point-like terms (which contain Dirac delta functions and their derivatives). This point-like terms are the contribution to the energy operator of the mean square radii: for each multipolarity order, there exists an infinite number of mean square radii, which contribute only to the overlap (contact) energy. Considering the contribution of the mean square radii is equivalent to considering the correct form of the bipolar expansion in the interpenetrating region. 
		
		As pointed out in Ref.\cite{cus}, the first consequence of using the correct form of the bipolar expansion in the overlap region is that the first order correction to the energy no longer vanishes (in Ref.\cite{mar}, where the contribution of the mean square radii have not been considered, the first-order correction to the interaction energy between two hydrogen atoms in their ground states is zero).

		The calculation of the first-order correction to the energy can be done either in the coordinate space (using the operator (\ref{oper})) or in the reciprocal (Fourier) space. We shall use here the second method. We take the Fourier transform $\widetilde{W}({\bf a}, {\bf b}; {\bf k})=\int d{\bf R} \;e^{i{\bf k}{\bf R}}\; W ({\bf a}, {\bf b}; {\bf R})$ and, after summing over the indices $n$, $n'$, we obtain:
		\end{multicols}
		\begin{align}
			\widetilde{W}({\bf a}, {\bf b}; {\bf k})= & \frac{4 \pi e^2}{k^2} \sum_{l,l',m,m'} (-1)^{l'} j_l(ka)j_{l'}(kb) Y_{lm}^* ( \hat{{\bf a}}) Y_{l'm'}^* ( \hat{{\bf b}}) Y_{lm} ( \hat{{\bf k}}) Y_{l'm'} ( \hat{{\bf k}})- \nonumber \\
			& \frac{4 \pi e^2}{k^2} \sum_{l,m} j_l(ka) Y_{lm}^* ( \hat{{\bf a}}) Y_{lm} ( \hat{{\bf k}}) - \\
			&  \frac{4 \pi e^2}{k^2} \sum_{l',m'} (-1)^{l'} j_{l'}(kb) Y_{l'm'}^* ( \hat{{\bf b}}) Y_{l'm'} ( \hat{{\bf k}}) + \frac{4\pi e^2}{k^2}. \nonumber
		\end{align}
		\begin{multicols}{2}
	\noindent	Due to the spherical symmetry of the ground-state wave function of the hydrogen atom, only the monopole mean square radii $(l=l'=m=m'=0)$ give contribution to the first-order correction to the energy:
	\end{multicols}	
		\begin{align}
			\widetilde{E}^{(1)}({\bf k})= \int d {\bf R}\;e^{i{\bf k} {\bf R}} E^{(1)}({\bf R})=  
			\int d{\bf a} \int d{\bf b} \; \Psi^{(0)*}({\bf a}) \Psi^{(0)*}({\bf b}) \widetilde{W}({\bf a}, {\bf b}; {\bf k}) \Psi^{(0)}({\bf a}) \Psi^{(0)}({\bf b}),
		\end{align}
	\begin{multicols}{2}	
		where \[ \Psi^{(0)}({\bf r}) = \frac{1}{\pi^{1/2} a_0^{3/2}} \; \exp \left(-\frac{r}{a_0} \right)\]
		is the hydrogen ground-state wave function, $a_0$ is the Bohr radius. All the integrations can be easily done and, after taking the inverse Fourier transform, one obtains:
		\begin{align}\label{cor1}
			E^{(1)}({\bf R})= \frac{E_0 a_0}{12R}\; \exp \left(- \frac{2R}{a_0} \right) \cdot \nonumber \\ \left(\frac{4R^3}{a_0^3} + \frac{18 R^2}{a_0^2} - \frac{15 R}{a_0} -24 \right), 
		\end{align}
		where $E_0= -e^2/(2a_0)$ is the ground-state energy of the hydrogen atom.
		This is the result obtained in Eq.(31) of Ref.\cite{pau}, by the direct integration of the Coulomb potential, when the exchange is neglected. As in our calculation this term have been obtained from the contribution of the monopole mean square radii, we conclude that it has nothing to do with the polarization of the atoms. The effect of the polarization (contributions from the dipole operator, quadrupole operator etc.) appears in the second order of the perturbation theory.

		\subsection*{\small{3.2 The interaction between a hydrogen atom and a proton}}
		\label{sec:32}
		The unperturbed system consists of a hydrogen atom in its fundamental state and a proton. We study their electrostatic interaction, neglecting the exchange (that is, we suppose that the electron does not jump from one proton to the other).
		The charge mean square radii of the two interacting systems are:
		\begin{align}
			\overline{r_{lm}^{2n}}^{(1)} = -e \sqrt{ \frac{4\pi}{2l+1}} a^{2n} Y_{lm}^*({\bf a}) + e \delta_{l,0} \delta_{n,0}, \nonumber \\
			 \overline{r_{l'm'}^{2n'}}^{(2)}= e \delta_{l',0} \delta_{n',0}
		\end{align}
		
		The perturbation operator is:
	
		\begin{align}
			W( {\bf a}; {\bf R})= \sum_{l,m,n} \frac{-4 \pi e^2 a^{2n}}{2^n n! (2l+2n+1)!!} \cdot \nonumber \\
			 Y_{lm}^*({\bf a}) \Delta_{R}^n Y_{lm}(-\nabla_{{\bf R}}) \frac{1}{R} + \frac{e^2}{R}
		\end{align}
		and its Fourier transform:
		\begin{align}
			\widetilde{W} ( {\bf a}; {\bf k})= - \frac{4\pi e^2 }{k^2} \sum_{l,m} j_l(ka) 
			Y_{lm}^*(\hat{\bf a}) Y_{lm}(\hat{\bf k}) \nonumber \\
			+\frac{4\pi e^2}{k^2}.
		\end{align}
		One obtains:
		
		\begin{align}
			\widetilde{E}^{(1)}({\bf k})=  \int d{\bf a} \psi^{(0)*}({\bf a}) \widetilde{W} ( {\bf a}; {\bf k}) \psi^{(0)}({\bf a})= \nonumber \\
			  \frac{4\pi e^2}{k^2} \left(1- \frac{16}{(4+k^2a_0^2)^2} \right),
		\end{align}
		and after taking the inverse Fourier transform we get the first-order correction to the energy:

		\begin{align}
			E^{(1)}({\bf R}) = \frac{e^2}{R} e^{ -2\frac{R}{a_0}} \left( \frac{R}{a_0} +1 \right)
		\end{align}
		This is the well-known result for the first-order correction to the energy, when the exchange is neglected (Eq. (19b) of Ref.\cite{pau}). As in the preceeding subsection, this correction to the energy comes from the contributions of the monopole mean square radii of the hydrogen atom, so it has nothing to do with the polarization.

		\section*{\small{4 Discussion and Conclusions}}
		\label{sec:4}

		Our main results (\ref{sf0}) - (\ref{carcon}), (\ref{potential}), (\ref{ensf})  contain series of the type $\sum_{l,m,n}[\dots] Y_{lm}(-\nabla) \Delta^n \frac{1}{R}= -4\pi \sum_{l,m,n}[\dots] Y_{lm}(-\nabla) \Delta^{n-1} \delta({\bf R})$.
		 Such series of Dirac delta functions were studied both by mathematicians and physicists, in connection with the asymptotic analysis, generalized solutions of differential equations,  multipole analysis, quantized field theories, etc. \cite{est, gel, row, seb1, seb2, lou, gord, gut}.
		 Starting from the need of multipole series in different branches of physics, the mathematicians extended the Schwartz's space of distributions to different spaces of ultradistributions (polynomial, tempered, with exponential growth, of compact support, Hermitean, etc), in which some  series of Dirac delta functions are convergent. Without entering into the mathematical details of these theories, we wish to underline here some important aspects regarding these distributional series:
		 \begin{itemize}
		\item our formalism make sense only for finite or rapidly descending charge densities, so that the mean square radii Eq.(\ref{raze}) should be finite. 
		 \item the distributional series which appear in our results are convergent at least as polynomial ultradistributions. Finding the largest space in which our results are convergent remains an open problem.
		 \item these infinite series of Dirac delta functions are not localized (supported) at the origin (\cite{row, gut}). If we regard them as functionals applied to some analytic function $\phi$, we obtain:
		 \begin{align} 
		 \sum_{n=0}^{\infty} c_n \delta^{(n)}({\bf R}) \phi({\bf R})= \sum_{n=0}^{\infty} (-1)^n c_n \phi^{(n)}(0),
		 \end{align}
		 so they depend not only on the values of $\phi$ at $0$, but on the values of $\phi$ in a certain neighborhood of $0$. In our multipole formalism these series are precisely connected to the finite or infinite overlap of the charge densities.
		 
		 \item 	When we calculate the interaction energy by using Eq. (\ref{enpot}), if the charge density $\rho_1({\bf r})$ is an analytic function, we obtain a convergent power series. Let us consider, for example, the interaction between a uniformly charged sphere $\rho_1({\bf r})= C_1 \theta(a-r)$ and a gaussian charge density $\rho_2({\bf r})= C_2 e^{-\lambda |{\bf r}- {\bf R}|^2}.$ 
		 From Eqs.(\ref{potential}), (\ref{enpot}), one obtains:
		 \begin{align}
		 W(R=0)= 8C_1 C_2 \frac{a^3 \pi}{3\lambda} - \frac{8 \pi^2 C_1 C_2}{15} a^5 + \dots ,
		 \end{align}
		 which are the first two terms of the Taylor series of the exact result:
		 \begin{align} 
		 W(R=0)= \frac{8 \pi^2 C_1 C_2}{3 \lambda} a^3 e^{- \lambda a^2} +  
		 \frac{4 \pi^2 C_1 C_2}{\lambda^{\frac{3}{2}}} a^2 \cdot \nonumber \\
		  \gamma \left( \frac{3}{2}, \lambda a^2 \right) -  \frac{4 \pi^2 C_1 C_2}{3 \lambda^{\frac{3}{2}}} a^2 \gamma \left( \frac{5}{2}, \lambda a^2 \right),
		 \end{align}
		 where $\gamma$ is the incomplete Gamma function.
		 
		 \item all the derivatives which appear in our formulas should be taken in distributional sense.
		  Let us consider  the interaction between a uniformly charged sphere  $\rho_1({\bf r})= C_1 \theta(a-r)$  and an exponential charge density  $\rho_2({\bf r}) = C_2 e^{-\lambda |{\bf r}-{\bf R}|}$. If we use Eq. (\ref{potential}) for the electrostatic potential of the sphere and calculate $ W=\int d {\bf r} \;\phi_1({\bf r}) \rho_2({\bf r}- {\bf R})$, when we integrate by parts we should apply the operator $\Delta^n$ to the exponential function, {\it without neglecting the Delta-type terms}:
		  \begin{align}
		  \Delta e^{-\lambda r} =  -2 \lambda \frac{e^{-\lambda r}}{r} + \lambda^2 e^{-\lambda r} \nonumber \\
		  \Delta^2 e^{-\lambda r} = - 4 \lambda^3 \frac{e^{-\lambda r}}{r} + \lambda^4 e^{-\lambda r} + 8 \pi \lambda \delta({\bf r}) \nonumber \\
		  \Delta^3 e^{-\lambda r} = -6 \lambda^5 \frac{e^{-\lambda r}}{r} + \lambda^6 e^{-\lambda r} + \nonumber \\16 \pi \lambda^3 \delta({\bf r}) + 8 \pi \lambda \Delta \delta({\bf r}),... \nonumber etc
		  \end{align}
		  For $n= \overline{0,\infty}$, we obtain a series of delta functions  and their derivatives, which should be summed.	We can do this by using the Fourier transform method:
		 
		 	\begin{align}
		 	W(R) = \int d {\vec r} \phi_1({\bf r}) \rho_2({\bf r})= \nonumber \\
		 	4 \pi C_1 C_2 \sum_{n=0}^{\infty} \frac{a^{2n+3}}{2^n n! (2n+3)!!} \int d{\bf r} \Delta^n \frac{1}{r} e^{- \lambda | {\bf r}- {\bf R}|}= \nonumber \\
		 	\frac{4}{3} \pi C_1 C_2 a^3 \int d{\bf r} 
		 	\frac{1}{r} e^{- \lambda | {\bf r}- {\bf R}|} - 16 \pi^2 C_1 C_2 \cdot \nonumber \\
		 	\sum_{n=1}^{\infty} \frac{a^{2n+3}}{2^n n! (2n+3)!!} \Delta_{\bf R}^{n-1}e^{-\lambda R}= \nonumber \\		 
		  \frac{4}{3} \pi C_1 C_2 a^3 \int d{\bf r} 
		 	\frac{1}{r} e^{- \lambda | {\bf r}- {\bf R}|}- 16 C_1 C_2 \lambda \cdot \nonumber \\
		 	\int d{\bf k} \frac{e^{-i {\bf k}{\bf R}}}{(k^2+\lambda^2)^2} \sum_{n=1}^{\infty} \frac{a^{2n+3} (-1)^{n-1} k^{2n-2}}{2^n n! (2n+3)!!}= \nonumber \\
		 	\frac{1}{r} e^{- \lambda | {\bf r}- {\bf R}|}+ \frac{16 C_1 C_2 \lambda^2}{\pi^2} \int d{\bf k} \frac{ e^{-i {\bf k}{\bf R}}}{k^3 (k^2+ \lambda^2)^2} \cdot \nonumber \\
		 	\left[ \frac{1}{4\pi i} j_1(ka) -\frac{ka}{3} \right]	
		 \end{align}
		 Note that the series which contain $a^{2n+3}$ from the above equation (lines 2, 4 and 6) should not be viewed as  power series in "a" (because their coefficients are generalized functions) and are not useful in numerical calculations.	
		 For $R=0$ one obtains:
		  \begin{align} \label{result}
		  W(R=0) =  \frac{16 \pi^2 C_1C_2}{\lambda^5} \cdot \nonumber \\ \left(e^{-\lambda a}(\lambda a+2)^2+ \lambda^2 a^2 -4 \right),
		  \end{align}
		  which coincide with the  result obtained by direct calculation.

		 \end{itemize}

		Using the multipolar formalism, as presented in Refs. \cite{dub, rad}, we have written four compact equivalent forms for the bipolar expansion of the inverse distance, which are valid in all the regions of configuration space. In the nonoverlapping region, these expansions coincide with those already known \cite{bue, sac, kay}. The equivalence of our results in the interpenetrating region with the other results from the literature remains an open problem. Using the first order perturbation theory, we have calculated the overlap corrections to the long-range electrostatic interaction energy for two simple systems, neglecting the exchange. We have shown that, considering the contributions of the mean square radii to the interaction energy is equivalent to considering the correct form of the bipolar expansion in the interpenetrating region.\\

\end{multicols}

	\end{document}